\documentclass[structabstract]{aa}
\usepackage{natbib}
\usepackage{graphicx}
\usepackage{textcomp}
\usepackage{longtable}
\usepackage{lscape}
\begin{document}
   \title{The FERRUM project: Transition probabilities for forbidden lines in [\ion{Fe}{ii}] and experimental metastable lifetimes}
\authorrunning{Gurell et al.}
\titlerunning{The FERRUM project: forbidden lines in \ion{Fe}{ii}}
%   \subtitle{I. Overviewing the $\kappa$-mechanism}

   \author{J. Gurell
          \inst{1}%\fnmsep\thanks{e-mail: jonas.gurell\@physto.se}
          \and
          H. Hartman\inst{2}
          \and
          R. Blackwell-Whitehead\inst{2}
          \and
          H. Nilsson\inst{2}
          \and
          E. B\"ackstr\"om\inst{1}
          \and
          L.O. Norlin\inst{3}
          \and
          P. Royen\inst{1}
          \and
          S. Mannervik\inst{1}
          }

   \institute{Department of Physics, Stockholm University, AlbaNova University Center, SE-10691 Stockholm, Sweden\\
              \email{jonas.gurell@physto.se}
         \and
             Lund Observatory, Lund University, Box 43, SE-22100 Lund, Sweden
             \and
                Department of Physics, Royal Institute of Technology, AlbaNova University Center, SE-10691 Stockholm, Sweden
             }

   \date{Received XXX; accepted XXX}

% \abstract{}{}{}{}{}
% 5 {} token are mandatory

  \abstract
   {Accurate transition probabilities for forbidden lines are important diagnostic parameters for low-density astrophysical plasmas. In this paper we present experimental atomic data for forbidden [\ion{Fe}{ii}] transitions that are observed as strong features in astrophysical spectra. }
   {To measure lifetimes for the $3d^6$($^3$G)$4s$\,a\,$^4$G$_{11/2}$ and $3d^6$($^3$D)$4s$\,b\,$^4$D$_{1/2}$ metastable levels in \ion{Fe}{ii} and experimental
transition probabilities for the forbidden transitions $3d^7$\,a\,$^4$F$_{7/2,9/2}$ -- $3d^6$($^3$G)$4s$\,a\,$^4$G$_{11/2}$.}
   {The lifetimes were measured at the ion storage ring facility CRYRING using a laser probing technique. Astrophysical branching fractions were obtained from spectra of Eta Carinae, obtained with the Space Telescope Imaging Spectrograph onboard the {\it Hubble Space Telescope}. The lifetimes and branching fractions were combined to yield absolute transition probabilities.}
   {The lifetimes of the a\,$^4$G$_{11/2}$ and the b\,$^4$D$_{1/2}$ levels have been measured and have the following values, $\tau=0.75\pm0.10$\,s and
   $\tau=0.54\pm0.03$\,s respectively. Furthermore, we have determined the transition probabilities for two forbidden transitions of a\,$^4$F$_{7/2,9/2}$ -- a\,$^4$G$_{11/2}$ at 4243.97 and 4346.85\,\AA . Both the lifetimes and the transition probabilities are compared to calculated values in the literature.}
   {}


   \maketitle
%
%________________________________________________________________

\section{Introduction}
The cosmic abundance of iron is relatively high compared to other iron group elements and the spectrum of singly ionized iron, \ion{Fe}{ii}, is a significant contributor to the spectral opacity of the sun and hotter stars. The complex energy level structure of \ion{Fe}{ii} makes the spectrum extremely line rich and it has been studied in great detail with more than 1000 energy levels identified in the literature (\citeauthor{Johansson09}\,\citeyear{Johansson09}).

\ion{Fe}{ii} lines are observed in the spectra of a wide variety of astronomical objects, and there is a considerable demand for accurate atomic
data for this ion. To meet the accurate data requirements of modern astrophysics, a program was initiated to supply the astronomical community with reliable atomic data: The FERRUM-project (\citeauthor{FERRUMproj}\,\citeyear{FERRUMproj}). The aim of this international collaboration is to measure and evaluate astrophysically relevant experimental and theoretical transition data for the iron group elements.

There are 62 metastable levels in \ion{Fe}{ii}. The parity forbidden lines from some of these levels are observed as prominent features in astrophysical low
density plasmas, such as nebulae, \ion{H}{ii} regions and circumstellar gas clouds. However, metastable levels have radiative lifetimes several orders of
magnitude longer than other levels and are thus more affected by collisions. Due to the absence of these lines in laboratory spectra, the majority of forbidden
line transition probabilities ($A$-values) available in the literature are from theoretical calculations.

There are only four metastable levels in \ion{Fe}{ii} with laboratory measured lifetimes. The a\,$^6$S$_{5/2}$ and b\,$^4$D$_{7/2}$ levels have been measured
by \citeauthor{Rostohar}\,(\citeyear{Rostohar}) using laser probing of a stored ion beam (a laser probing technique, LPT). In addition, the a\,$^4$G$_{9/2}$
and b\,$^2$H$_{11/2}$ levels have been measured by \citeauthor{Hartman}\,(\citeyear{Hartman}) using the LPT. There is good agreement between
\citeauthor{Rostohar}\,(\citeyear{Rostohar}) and the calculated values of \citeauthor{Nussbaumer}\,(\citeyear{Nussbaumer}) and
\citeauthor{Quinet}\,(\citeyear{Quinet}). However, the lifetimes of \citeauthor{Rostohar}\,(\citeyear{Rostohar}) are systematically shorter than the
calculations of Garstang (\citeyear{Garstang}). \citeauthor{Hartman}\,(\citeyear{Hartman}) also combined the lifetimes of a\,$^6$S$_{5/2}$, b\,$^4$D$_{7/2}$
and a\,$^4$G$_{9/2}$ with branching fractions ($BF$s) to determine experimental $A$-values for forbidden transitions. \citeauthor{Hartman}\,\citeyear{Hartman}
measured the $BF$s in astrophysical spectra observed in the ejecta of Eta Carinae and presented additional theoretical $A$-value calculations.

We present radiative lifetimes for the a\,$^4$G$_{11/2}$ and b\,$^4$D$_{1/2}$ metastable levels in \ion{Fe}{ii} measured using the LPT at the CRYRING facility.
In addition, $BF$s for two forbidden transitions 4243.97 \AA\ and 4346.85 \AA\ (a$^4$F$_{9/2}$ - a$^4$G$_{11/2}$ and a$^4$F$_{7/2}$ - a$^4$G$_{11/2}$) have
been measured in astrophysical spectra observed in the ejecta of Eta Carinae recorded with the {\it Hubble Space Telescope} ({\it HST}) Space Telescope Imaging
Spectrograph (STIS). The radiative lifetimes have been combined with the $BF$s to yield $A$-values and we provide a comparison with theoretical values in the
literature.

%Six metastable levels have been measured in \ion{Fe}{ii}. There is a good agreement between experimental and calculated values with a preference for the
%calculations by \citeauthor{Quinet} (\citeyear{Quinet}). The largest discrepancy is for the b\,$^2$H$_{11/2}$ level where the experimental value is
%$\tau=3.8\pm0.3$\,s  and the theoretical values (\citeauthor{Quinet}\,\citeyear{Quinet}) are 6.59\,s and 5.20\,s from the Superstructure
%code(\citeauthor{Eissner}\,\citeyear{Eissner};\citeauthor{NussbaumerSST}\, \citeyear{NussbaumerSST}) and the relativistic Hartree-Fock (HFR) code
%respectievely(\citeauthor{Cowan}\,\citeyear{Cowan}). The discrepancy was explained by \citeauthor{Hartman} (\citeyear{Hartman}) through a level mixing between
%a\,$^4$G$_{11/2}$ and b\,$^2$H$_{11/2}$ which was not reproduced by the calculations. This level mixing had also been observed in earlier work when studying
%allowed transitions in \ion{Fe}{ii} (\citeauthor{Johansson}\,\citeyear{Johansson}). We have measured the lifetimes of the a\,$^4$G$_{11/2}$ and
%b\,$^4$D$_{1/2}$ levels to gain an additional understanding of this level mixing and the results are presented in this paper.

\section{Experimental Measurements}

The lifetime ($\tau$) of a level $i$ is defined as the inverse of the total transition probability ($A_{ik}$) for all possible decay channels from that level,
according to Eq.~\ref{lifetime}, where $k$ is summed over all lower levels.

\begin{equation}
\tau_i=\frac{1}{\sum_kA_{ik}}\label{lifetime}
\end{equation}

The majority of all excited levels will decay through electric dipole ({\it E1}) transitions which typically gives a lifetime of the order of a nanosecond, but
if no allowed {\it E1} decay channels are available, i.e. the level is metastable, the population decay has to be through higher order transitions associated
with magnetic dipole or electric quadrupole terms. The probability for such a transition to occur is several orders of magnitude smaller compared to the
probability for an {\it E1} transition which makes the lifetime of a metastable level significantly longer, typically several milliseconds, seconds, minutes or
even hours.

The $BF$ of a transition $ik$ is defined as the transition probability, $A_{ik}$, for a single line divided by the sum of the transition probabilities from all lines with the
common upper level. according to Eq.~\ref{BF}.

\begin{equation}
BF_{ik}=\frac{A_{ik}}{\sum_kA_{ik}}\label{BF}
\end{equation}

By combining Eqs.~\ref{lifetime} and~\ref{BF} the absolute transition probability, referred to as the $A$-value, of each decay channel can be deduced according to Eq.~\ref{A}.

\begin{equation}
A_{ik}=\frac{BF_{ik}}{\tau_i}\label{A}
\end{equation}

\subsection{Lifetime measurements of metastable \ion{Fe}{II} levels}

The lifetimes of the two metastable levels were measured using the LPT (\citeauthor{Lidberg}\,\citeyear{Lidberg};
\citeauthor{Mannervik2003}\,\citeyear{Mannervik2003}; \citeauthor{Mannervik2005}\,\citeyear{Mannervik2005}) at the ion storage ring CRYRING at the Manne
Siegbahn laboratory in Stockholm, Sweden. The LPT has been developed and refined during several years and has been used to measure lifetimes ranging from 3.4
ms in \ion{Xe}{ii} (\citeauthor{LidbergXe}\,\citeyear{LidbergXe}) to 89\,s in \ion{Ba}{ii} (\citeauthor{Gurell}\,\citeyear{Gurell}).

A hot ion source is used to produce ions that can be extracted and injected into the storage ring. Once inside CRYRING, singly charged ions
may be stored for several minutes at 40\,keV and investigated in this ultra high vacuum environment with a background pressure lower than $10^{-11}$\,Torr.
For the experiment described in this paper Fe$^+$ ions were produced in a Nielsen ion source used in combination with an oven filled with FeCl$_2$. By heating the oven to approximately 400$^\circ$\,C and running the ion source with argon as a carrier gas it was possible to extract and store a current of Fe$^+$ ions as high as 4\,$\mu$A in CRYRING.

\begin{figure}
\centering
\includegraphics[width=9cm]{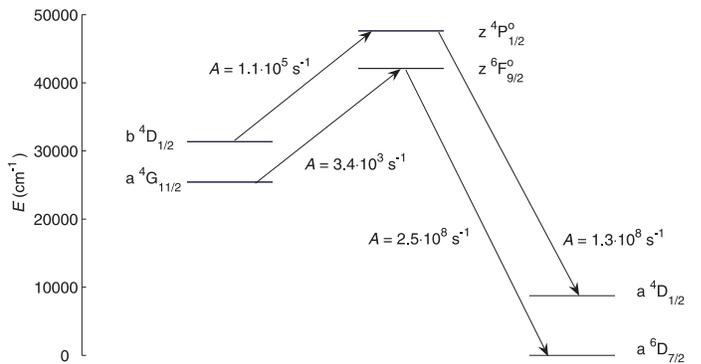}
\caption{Probing scheme for the two levels studied. The populations of the metastable levels, shown in the left side of the figure, were probed through
excitation to the higher lying levels in the middle of figure. Once in the higher lying excited levels the populations decay to lower lying levels, shown in
the right side of the figure. The intensity of this fluorescence is detected and used as a relative measurement of the original population of the metastable
level under study. See the text for references to the cited $A$-values.} \label{scheme}
\end{figure}

Ions in excited states will decay as they are being stored and by passively monitoring the excited level population, a decay curve can be constructed.
However, the low intensity of forbidden transitions is hard to detect directly since the emission is spread out over the 52\,m length of the storage ring and
spread across the 4$\pi$ solid angle. A LPT in combination with a Doppler tuning device (DTD) was used to increase the strength of the signal. The LPT probes
the population of the metastable level by laser--inducing a transition to a higher lying short lived level. Within a few ns this level will decay to another
lower lying level and the fluorescence originating from that transition is used as an indirect measurement of the population of the metastable level at the
time of the probe pulse's arrival. The probing scheme used in this study is shown in Fig.~\ref{scheme} with $A$-values from \citeauthor{Baschek}
(\citeyear{Baschek}), \citeauthor{Kostyk} (\citeyear{Kostyk}) and \citeauthor{ASD} (\citeyear{ASD}) and energy levels from \citeauthor{Johansson}
(\citeyear{Johansson}).

The DTD was used to locally accelerate the ions in front of our detector, thereby inducing a Doppler shift. A narrow band ring dye laser was used to tune the resonance of the ions within the small volume of a few cm$^{3}$ that constitutes the DTD. This technique forces the probing as
well as the following rapid fluorescence to occur in front of the detector which greatly enhances the photon count.

We used a Coherent 699-29 ring dye laser, operated with a Rhodamine 6G dye and pumped by a Coherent Innova 400-25 argon laser, to produce laser
light with the desired wavelengths of 5985 and 6143\,\AA. The fluorescence was monitored with a Hamamatsu R585 photomultiplier tube in front of which an
optical filter was mounted in order to reduce the background signal coming from scattered laser light.

The populations of the metastable levels under study were probed as a function of delay time after ion injection. Every fourth ion injection was used to
measure the laser induced fluorescence at a fixed time after injection in order to make sure that the initial population of the level of interest did not
change during the data acquisition time.

\subsubsection{Collisional quenching}
The small amount of residual gas in the storage ring gives rise to collisional quenching of the metastable levels. Therefore, it is necessary to measure the
decay rate of the level population as a function of pressure and extrapolate to zero pressure in order to get the pure radiative decay rate. The decay rate has a
linear dependence and the extrapolation can be determined with a Stern-Vollmer plot (\citeauthor{Demtroder}\,\citeyear{Demtroder}). In practice this is achieved by allowing the residual gas pressure inside CRYRING to be varied by heating a non-evaporative getter pump, which releases previously absorbed rest gas thereby increasing the pressure slightly. The operating pressure inside CRYRING is out of the range of standard vacuum meters and a relative pressure measurement has to be used instead. Once stored, ions are gradually neutralized due to collisions with the rest gas and the number of ions stored undergoes an exponential decay as a function of time after injection. This decay rate of the ion beam is assumed to be inversely proportional to the rest gas pressure and is used as a relative pressure measurement. The decay of the ion beam is measured with a multi-channel plate connected to the CRYRING storage ring.

\begin{figure}
   \centering
   \includegraphics[width=8cm]{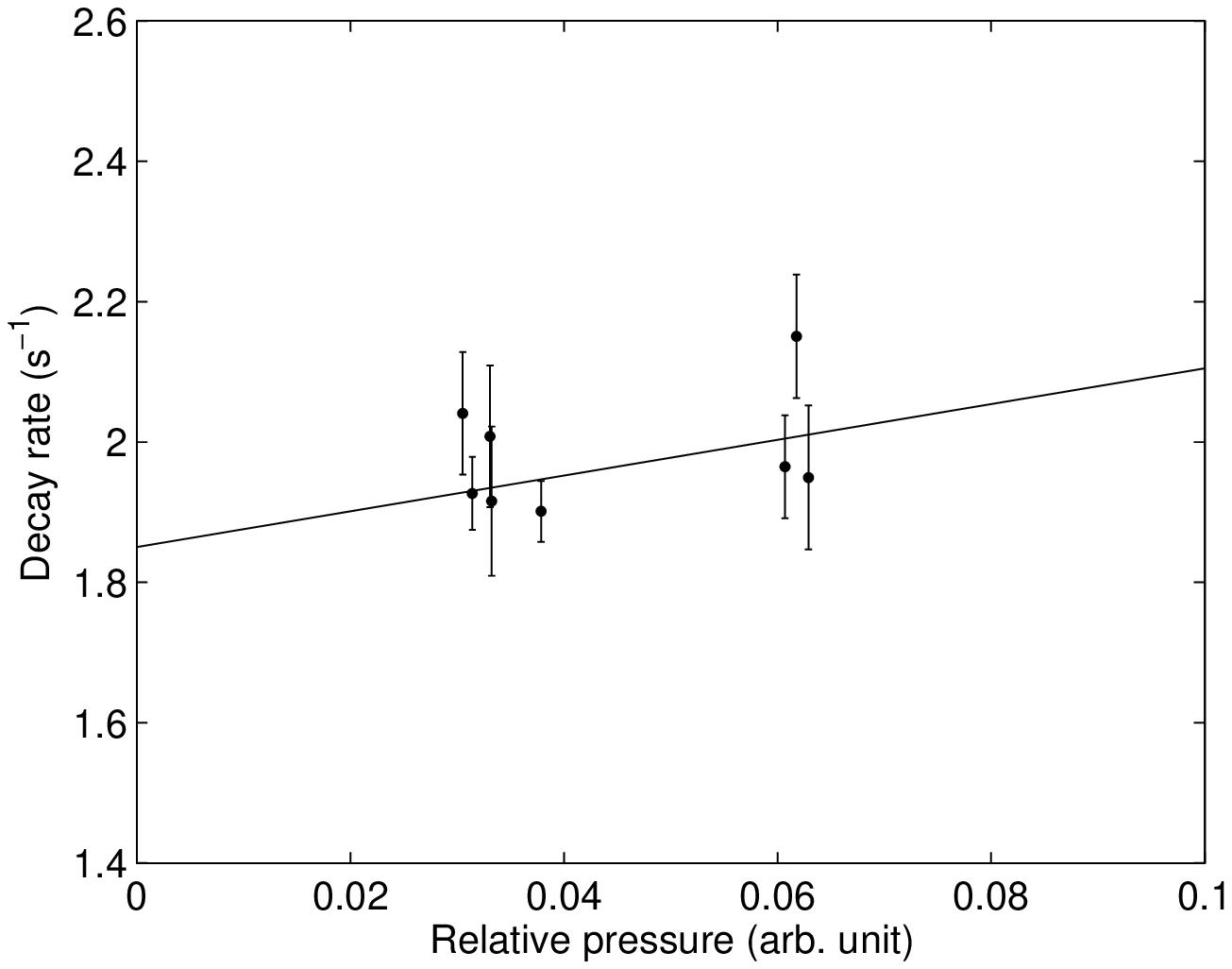}
      \caption{Stern-Vollmer plot showing the decay rate of the $3d^6$($^3$D)$4s$\,b$^4$D$_{1/2}$ level population as a function of pressure.} \label{SV4D}
   \end{figure}

\begin{figure}
   \centering
   \includegraphics[width=8cm]{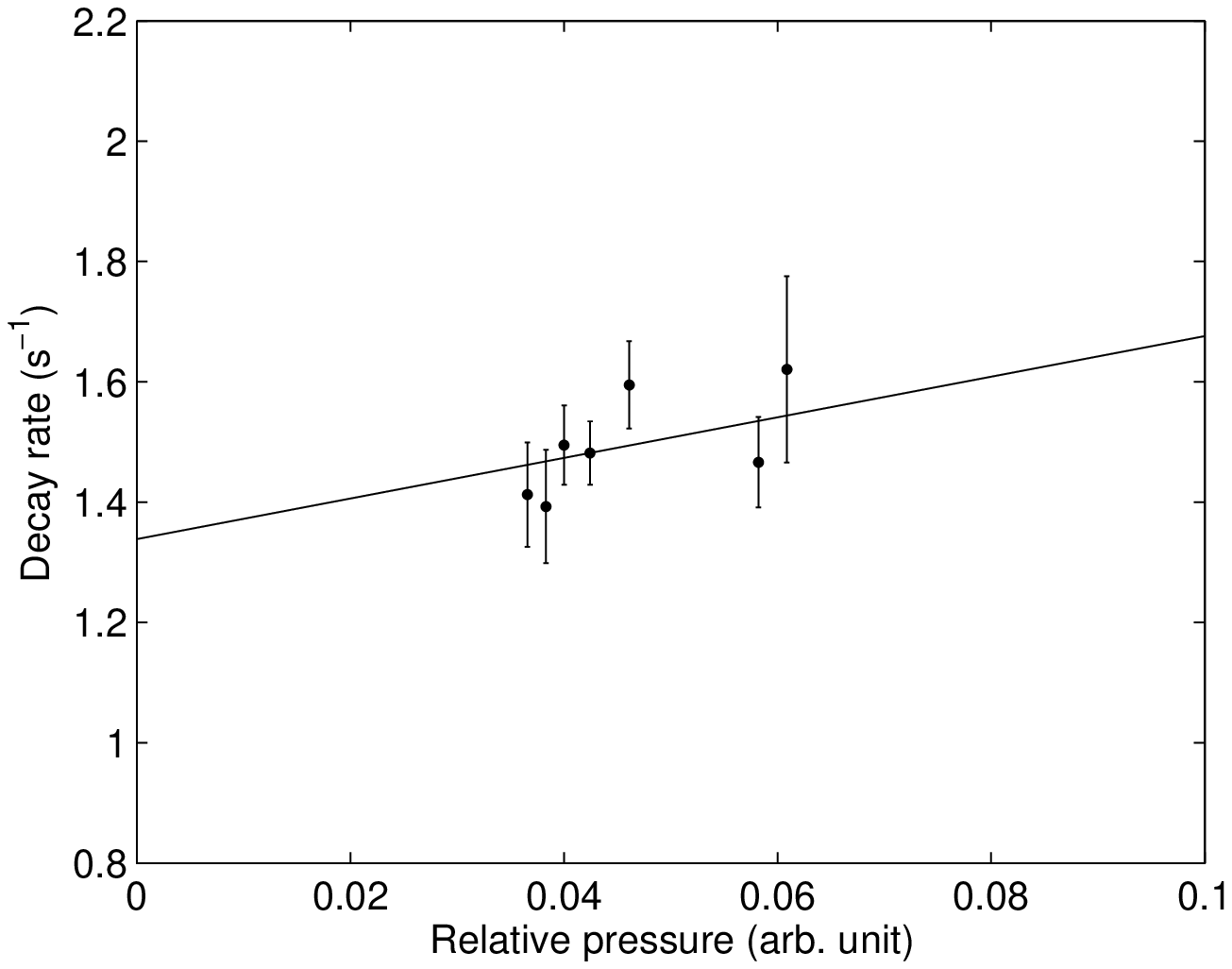}
      \caption{Stern-Vollmer plot showing the decay rate of the $3d^6$($^3$G)$4s$\,a$^4$G$_{11/2}$ level population as a function of pressure.} \label{SV4G}
   \end{figure}

The extrapolations of the decay rates to zero pressure were made in two Stern-Vollmer plots shown in Figs.~\ref{SV4D} and \ref{SV4G}. The data were fitted with
a linear fit weighted against the inverse uncertainty in each data point. The uncertainty of the extrapolated pure radiative decay rate is given as the
standard deviation of the fit in the corresponding Stern-Vollmer plot.

\subsubsection{Repopulation}
\begin{figure}
   \centering
   \includegraphics[width=8cm]{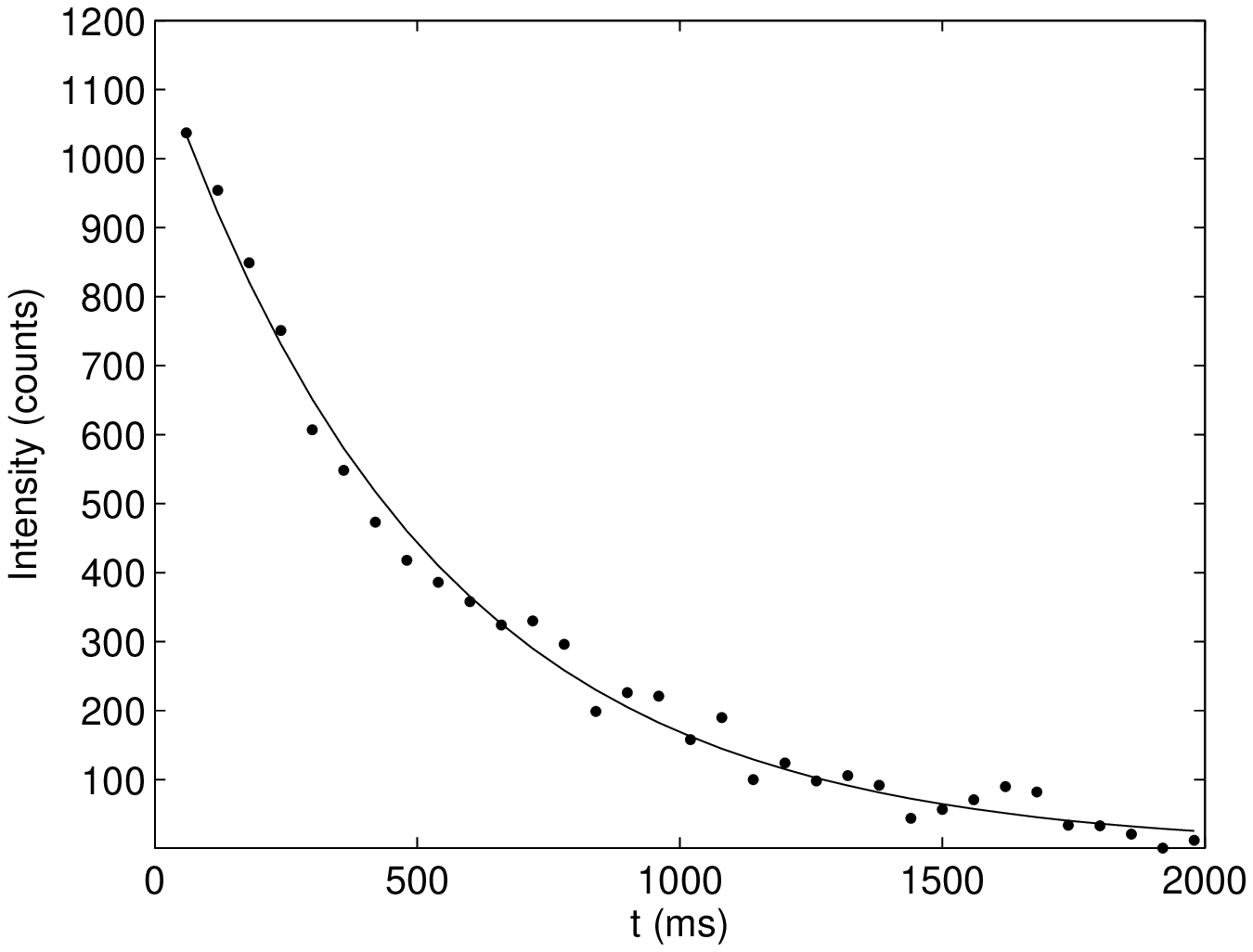}
      \caption{Observed population decay curve of the $3d^6$($^3$D)$4s$\,b\,$^4$D$_{1/2}$ level.} \label{LIF4D}
   \end{figure}

\begin{figure}
   \centering
   \includegraphics[width=8cm]{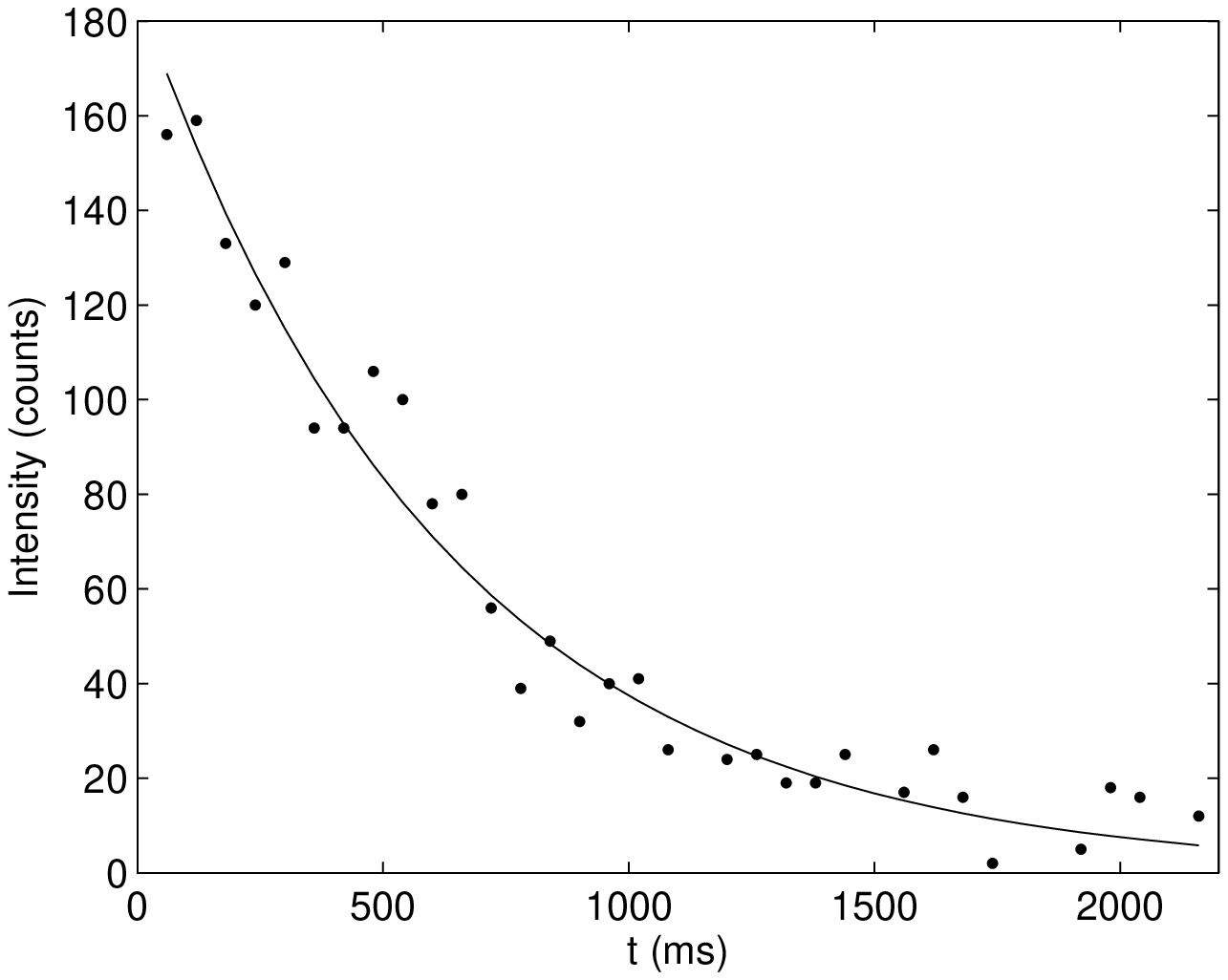}
      \caption{Observed population decay curve of the $3d^6$($^3$G)$4s$\,a\,$^4$G$_{11/2}$ level.} \label{LIF4G}
   \end{figure}

Collisional excitation of ground state ions into metastable levels, referred to as repopulation, is usually corrected for when using the LPT for lifetime
measurements, see e.g. \citeauthor{Royen} (\citeyear{Royen}). An attempt to measure the repopulation in this study was made but the signal was below the
background level of approximately 5 photons per second which makes the collisional excitation effect negligible. The production of metastable ions at the time
of ion injection was monitored and showed stable conditions throughout the experiment. Decay curves for the populations of the two levels are shown in
Figs.~\ref{LIF4D} and \ref{LIF4G} together with the fitted curves used to determine the lifetimes of the levels. The standard deviation of each fit was used as
the uncertainty in the particular lifetime curve.

\subsection{Astrophysical branching fractions for a\,$^4$\emph{G}$_{11/2}$}

The intensity of an emission line is proportional to the product $NA$ where $N$ is the population of the upper level and $A$ is the transition rate of the
transition. For lines having the same upper level the observed photon intensity is proportional to the transition rate. Together with Eq. 2 this relates the
$BF$ to the measured intensities for all the decay channels from a single upper level:

\begin{equation}
BF_{ik}=\frac{A_{ik}}{\sum_kA_{ik}}=\frac{I_{ik}}{\sum_kI_{ik}}  \label{BF2}
\end{equation}

For electric dipole transitions, the branching fractions are often determined using a standard laboratory light source with a high-density plasma and high
resolution spectrometer. However, the $A$-value of a forbidden line is typically $10^6-10^9$ times lower than the $A$-value of an allowed transition which
makes the forbidden lines too weak to be observed in the spectrum of a standard laboratory source. To observe strong forbidden lines the ions must be kept in a
low-density plasma environment, to minimize the collisional deexcitation (quenching), and in a large volume, to get a critical number of ions. Low density
astrophysical plasmas fulfil these criteria, and their spectra often show strong forbidden lines. In particular, the spectra of the Weigelt blobs in Eta
Carinae  are rich in forbidden transition from the iron group including the [\ion{Fe}{ii}] 4243 and 4346 \AA\ lines and Weigelt blob spectra observed by HST
STIS have been used in this study. A great advantage of using HST STIS is the high angular resolution (0$\farcs$1) which minimizes the contribution from
surrounding regions. In addition, the spectrum of the Weigelt blobs has been studied in great detail (e.g. \citeauthor{zphd01}\,\citeyear{zphd01}), which
decreases the uncertainty from blending lines or misinterpretations of spectral features.

Astrophysical $BF$s have similar uncertainties to laboratory measurements including uncertainties in the intensity measurements, intensity calibration and
unobservable lines (\citeauthor{Sikstrom02}\,\citeyear{Sikstrom02}). Additional uncertainties include corrections for interstellar reddening and blending from
other spatial regions along the line of sight. Furthermore, a standard reddening curve was used to correct for redding. However, the redding effect was found
to be only a few percent because the lines are separated by a relative narrow wavelength range of 100 \AA . The uncertainties have been treated in accordance
with the guidelines the National Institute of Standards and Technology, USA (\citeauthor{Taylor94}\,\citeyear{Taylor94}).

\begin{table*}
\caption{Experimentally determined lifetimes of metastable levels in \ion{Fe}{ii}. The uncertainties are given in brackets.} \label{Lifetimes} \centering
\begin{tabular}{@{}llcccccccc}     % 7 columns
\hline\hline
& & \multicolumn{8}{c}{$\tau$ (s)} \\
Configuration & Term &  This work & Ref. [1] & Ref. [2] & Ref. [3] & Ref. [4] & Ref. [5]& Ref. [6] & Ref. [7]\\
\hline
 $3d^54s^2$  & a\,$^6$S$_{5/2}$ &  & 0.23(3) & & 0.326 & 0.235 & 0.262 & 0.220 &  0.222\\
 $3d^6$($^3$G)$4s$ & a\,$^4$G$_{11/2}$ & 0.75(10) & & & 0.852 &  & 0.774 & 0.704 & \\
 $3d^6$($^3$G)$4s$ & a\,$^4$G$_{9/2}$ &  & & 0.65(2) & 0.856 &  & 0.755 & 0.694 &\\
 $3d^6$($^3$H)$4s$ & b\,$^2$H$_{11/2}$ &  & & 3.8(3) & 10.1 &  & 6.59 & 5.20 & \\
 $3d^6$($^3$D)$4s$ & b\,$^4$D$_{1/2}$ & 0.54(3) & & & 0.677 & & 0.630 & 0.550 &\\
 $3d^6$($^3$D)$4s$ & b\,$^4$D$_{7/2}$ &  & 0.53(3) & & 0.618 & & 0.567 & 0.500 & \\
\hline
References: & \multicolumn{6}{l}{[1] \citeauthor{Rostohar} (\citeyear{Rostohar}) (Experimental)}\\
& \multicolumn{6}{l}{[2] \citeauthor{Hartman} (\citeyear{Hartman}) (Experimental)}\\
& \multicolumn{6}{l}{[3] \citeauthor{Garstang} (\citeyear{Garstang}) (Calculated)}\\
& \multicolumn{6}{l}{[4] \citeauthor{Nussbaumer} (\citeyear{Nussbaumer}) (Calculated, Superstructure code)}\\
& \multicolumn{6}{l}{[5] \citeauthor{Quinet} (\citeyear{Quinet}) (Calculated, Superstructure)}\\
& \multicolumn{6}{l}{[6] \citeauthor{Quinet} (\citeyear{Quinet}) (Calculated, relativistic Hartree-Fock)}\\
& \multicolumn{6}{l}{[7] \citeauthor{Hartman} (\citeyear{Hartman}) (Calculated, CIV3 code)}\\
\hline
\end{tabular}
\end{table*}

\begin{table*}
\caption{Experimentally determined transition probabilities compared to theoretical values from a\,$^4$G$_{11/2}$.} \label{BFs} \centering
\begin{tabular}{@{}lllll}     % 7 columns
\hline\hline
\multicolumn{2}{l}{Transition} & \multicolumn{3}{l}{$A$ (s$^{-1}$)} \\
 & $\lambda_{air}$ (\AA) & This work & \multicolumn{2}{c}{\citeauthor{Quinet} (\citeyear{Quinet} )} \\
  & & & SST & HFR \\
\hline
a$^4$F$_{9/2}$ - a$^4$G$_{11/2}$ & 4243.97 & 1.05(15) & 1.02 & 1.12 \\
a$^4$F$_{7/2}$ - a$^4$G$_{11/2}$ & 4346.85 & 0.25(5) & 0.23 & 0.25 \\
\hline
\end{tabular}
\end{table*}

\section{Results}

A summary of the lifetime measurements and a comparison to theoretical values in the literature is given in Table~\ref{lifetime}. For the
$3d^6$($^3$G)$4s$\,a\,$^4$G$_{11/2}$ level $\tau=0.75\pm0.10$\,s  and for the $3d^6$($^3$D)$4s$\,b\,$^4$D$_{1/2}$ level $\tau=0.54\pm0.03$\,s. Theoretical
calculations indicate that unobserved branching transitions for the a\,$^4$G$_{11/2}$ level constitute less than 4\% of the total $BF$
(\citeauthor{Quinet}\,\citeyear{Quinet}). The large relative uncertainty associated with the a\,$^4$G$_{11/2}$ lifetime is due to a lower number of photon
counts per second compared to the detection of fluorescence following excitation from the b\,$^4$D$_{1/2}$ level. This is mainly due to the difference in the
$A$-values of the two probing transitions used, see Fig.~\ref{scheme}, which differ by almost two orders of magnitude. The experimental conditions not
associated with the internal structure of the atomic system (i.e. laser power, background pressure, ion current, optical alignment etc.) remained very similar
during the two lifetime measurements.

We have measured astrophysical $BF$s for lines from a\,$^4$G$_{11/2}$, from which the strongest decay channels are to a\,$^4$F at 4243 and 4346 \AA . The
transitions from b\,$^4$D$_{1/2}$ are distributed over a larger wavelength region than transitions from a\,$^4$G$_{11/2}$, and many of the lines are blended
with other lines making the intensity measurements uncertain. Therefore we could not derive $BF$s for lines from this level. The experimental $A$-values are
given in Table~\ref{BFs}, along with uncertainties and comparisons to calculations. The agreement with the values by \citeauthor{Quinet}\,(\citeyear{Quinet})
is within our uncertainties. We estimate the uncertainty of the $BF$s to 6-12\%, including the uncertainties in the intensity measurement, calibration,
reddening and residual lines.

\section{Discussion}

The relativistic Hartree-Fock (HFR) calculations by \citeauthor{Quinet}\,(\citeyear{Quinet}) show the best general agreement with the experimental values
presented in this and previous studies. The agreement is good with the exception of the b\,$^2$H$_{11/2}$ level (\citeauthor{Rostohar}\,\citeyear{Rostohar})
which differs by $5\sigma$. This deviation has been explained by \citeauthor{Hartman}\,\citeyear{Hartman} as an effect of level mixing between the
b\,$^2$H$_{11/2}$ and the a\,$^4$G$_{11/2}$ levels. In addition, this mixing has been observed by \citeauthor{Johansson}\,(\citeyear{Johansson}) through
unexpected spin forbidden spectral lines, e.g. the b\,$^2$H$_{11/2}$ - z\,$^6$F$_{9/2}$ transition at 6269.97 \AA .

Comparison of the theoretical lifetime values in Table~\ref{Lifetimes} for a\,$^4$G$_{11/2}$ and b\,$^2$H$_{11/2}$ indicate that the lifetime of the
a\,$^4$G$_{11/2}$ level should be approximately one tenth of the lifetime of b\,$^2$H$_{11/2}$. However, the experimental values reveal that the
a\,$^4$G$_{11/2}$ level is only one fifth of the lifetime for b\,$^2$H$_{11/2}$ indicating that the mixing effect can be observed in the lifetimes.
Calculations are sensitive to level mixing and performing {\it ab initio} calculations which reproduce the mixing properties of the system is not trivial. This
is illustrated in \citeauthor{Nahar} (\citeyear{Nahar}), \citeauthor{Nahar2} (\citeyear{Nahar2}), \citeauthor{HibbertFe} (\citeyear{HibbertFe}),
\citeauthor{Correge} (\citeyear{Correge}) and \citeauthor{Pickering_mixing} (\citeyear{Pickering_mixing}). In particular, \citeauthor{Pickering_mixing}
(\citeyear{Pickering_mixing}) emphasized that even though the agreement between the theoretical and experimental \ion{Fe}{ii} $A$-values investigated
\citeauthor{Pickering_mixing} (\citeyear{Pickering_mixing}) are in general extremely good, probabilities related to spin forbidden transitions that occur due
to level mixing may differ by up to an order of magnitude.

%We have measured radiative lifetimes for two metastable levels in [\ion{Fe}{ii}], and $BF$s and $A$-values for two forbidden transitions. In total there are %only six of a possible 62 metastable levels in [\ion{Fe}{ii}] with experimental radiative lifetimes reported in the literature. In general the theoretical %calculations have been shown to agree with experimental values where there is only weak level mixing. However, the calculated values can deviate significantly %if strong level mixing is present. We encourage further theoretical calculations to compliment future experimental measurements and astrophysical observations.

\begin{acknowledgements}
The authors acknowledge the help from the CRYRING staff. This work was supported by the Swedish Research Council (VR), the Swedish National Space Board (SNSB)
and through a Linnaeus grant from VR. RBW would like to acknowledge the European Commission for a Marie Curie Intra-European fellowship. This research has made use of the data archive for the HST Treasury Program on Eta Carinae (GO 9973) which is available online at \texttt{http://etacar.umn.edu}. The archive is supported by the University of Minnesota and the Space Telescope Science Institute under contract with NASA. In addition, we gratefully acknowledge the discussion with Prof. Selvelli on forbidden iron lines in RR Tel.\end{acknowledgements}


\begin{thebibliography}{}

\bibitem[\protect\citeauthoryear{Baschek et al.}{1970}]{Baschek}
Baschek, B., Garz, T., Holweger, H. \& Richter J. 1970, \aap, 4, 229

\bibitem[\protect\citeauthoryear{Corr\'eg\'e \& Hibbert}{2006}]{Correge}
Corr\'eg\'e, G. \& Hibbert, A. 2006, \apj, 636, 1166

%\bibitem[\protect\citeauthoryear{Cowan}{1981}]{Cowan}
%Cowan, R. D. 1981, The Theory of Atomic Structure and Spectra (Berkeley Univ. of California Press)

\bibitem[\protect\citeauthoryear{Demtr\"oder}{1996}]{Demtroder}
Demtr\"oder, W. 1996, Laser spectroscopy. Basic concepts and instrumentation, 2nd Ed. (New York, Berlin: Springer)

%\bibitem[\protect\citeauthoryear{Eissner et al.}{1974}]{Eissner}
%Eissner, W., Jones, M. \& Nussbaumer, H. 1974, Comput. Phys. Commun., 8, 270

\bibitem[\protect\citeauthoryear{Garstang}{1962}]{Garstang}
Garstang, R.H. 1962, \mnras, 124, 321

\bibitem[\protect\citeauthoryear{Gurell et al.}{2007}]{Gurell}
Gurell, J. et al. 2007, \pra, 75, 052506

\bibitem[\protect\citeauthoryear{Hartman et al.}{2003}]{Hartman}
Hartman, H. et al. 2003, \aap, 397, 1143

\bibitem[\protect\citeauthoryear{Hibbert \& Corr\'eg\'e}{2005}]{HibbertFe}
Hibbert, A. \& Corr\'eg\'e, G. 2005, \physscr, T119, 61

\bibitem[\protect\citeauthoryear{Johansson}{1978}]{Johansson}
Johansson, S. 1978, \physscr, 18, 217

\bibitem[\protect\citeauthoryear{Johansson et al.}{2002}]{FERRUMproj}
Johansson, S. et al. 2002, \physscr, T100, 71

\bibitem[\protect\citeauthoryear{Johansson}{2009}]{Johansson09}
Johansson, S. 2009, \physscr, T143, 014013

\bibitem[\protect\citeauthoryear{Kostyk \& Orlova}{1982}]{Kostyk}
Kostyk, R.I. \& Orlova, T.V. 1982, Astrometiya Astrofiz., 47, 32

\bibitem[\protect\citeauthoryear{Lidberg et al.}{1997}]{LidbergXe}
Lidberg, J., Al-Khalili, A., Cowan, R.D., Norlin, L.O., Royen, P. \& Mannervik, S. 1997, \pra, 56, 4

\bibitem[\protect\citeauthoryear{Lidberg et al.}{1999}]{Lidberg}
Lidberg, J., Al-Khalili, A., Norlin, L.O., Royen, P., Tordoir, X. \& Mannervik, S. 1999, Nucl. Instrum. Methods Phys. Res. B, 152, 157

\bibitem[\protect\citeauthoryear{Mannervik}{2003}]{Mannervik2003}
Mannervik, S. 2003, \physscr, T119, 49

\bibitem[\protect\citeauthoryear{Mannervik et al.}{2005}]{Mannervik2005}
Mannervik, S., Ellmann, A., Lundin, P., Norlin, L.O., Rostohar, D., Royen P. \& Schef, P. 2005, \physscr, T105, 67

\bibitem[\protect\citeauthoryear{Nahar \& Pradhan}{1994}]{Nahar}
Nahar, S.N. \& Pradhan, A.K. 1994, J. Phys. B, 27, 429

\bibitem[\protect\citeauthoryear{Nahar}{1995}]{Nahar2}
Nahar, S.N. 1995, \aap, 293, 967

%\bibitem[\protect\citeauthoryear{Nielsen et al.}{2009}]{Nielsen09}
%Nielsen, K.E., Kober, G.V., Weis, K. Gull, T.R., Stahl, O., Bomans, D.J. 2009, \apjs, 181, 473

%\bibitem[\protect\citeauthoryear{Nussbaumer \& Storey}{1978}]{NussbaumerSST}
%Nussbaumer, H. \& Storey, P.J. 1978, \aap, 64, 139

\bibitem[\protect\citeauthoryear{Nussbaumer et al.}{1981}]{Nussbaumer}
Nussbaumer, H., Pettini, M. \& Storey, P.J. 1981, \aap, 102, 351

\bibitem[\protect\citeauthoryear{Pickering et al.}{2002}]{Pickering_mixing}
Pickering, J.C., Donnelly, M.P., Nilsson, H., Hibbert, A. \& Johansson, S. 2002, \aap, 396, 715

\bibitem[\protect\citeauthoryear{Quinet et al.}{1996}]{Quinet}
Quinet, P., Le Dourneuf, M. \& Zeippen, C.J. 1996, \aaps, 120, 361

\bibitem[\protect\citeauthoryear{Ralchenko et al.}{2008}]{ASD}
Ralchenko, Yu., Kramida, A.E., Reader, J. \& NIST ASD Team (2008) NIST Atomic Spectra Database (version 3.1.5), [Online]. Available:
http://physics.nist.gov/asd3 [2008, September 2]. National Institute of Standards and Technology, Gaithersburg, MD.

\bibitem[\protect\citeauthoryear{Rostohar et al.}{2001}]{Rostohar}
Rostohar, D., Derkatch, A., Hartman, H., Johansson, S., Lundberg, H., Mannervik, S., Norlin, L.O., Royen, P. \& Schmitt, A. 2001, \prl, 86, 8

\bibitem[\protect\citeauthoryear{Royen et al.}{2007}]{Royen}
Royen, P., Gurell, J., Lundin, P., Norlin, L.O. \& Mannervik, S. 2007, \pra, 76, 030502(R)

\bibitem[\protect\citeauthoryear{Sikstr\"{o}m et al.}{2002}]{Sikstrom02} Sikstr\"{o}m C.M., Nilsson H., Litzen U., Blom A., Lundberg H., 2002, J. Quant. Spec. Rad. Trans., 74, 355

\bibitem[\protect\citeauthoryear{Taylor \& Kuyatt}{1994}]{Taylor94}
Taylor, B.N., Kuyatt, C.E. 1994, NIST Technical Note 1297, Guidelines for evaluating and expressing the uncertainty of NIST measurement results,
National Institute of Standards and Technology, USA (http://physics.nist.gov/Pubs/guidelines/).

\bibitem[\protect\citeauthoryear{Zethson}{2001}]{zphd01}
Zethson, T. 2001, PhD thesis, Lund University

\end{thebibliography}
\end{document}